# Two-step melting of $Na_{41}^+$

Sébastien Zamith*, Pierre Labastie*, Fabien Chirot**, and Jean-Marc L'Hermite*,

**LCAR/IRSAMC, Université de Toulouse-UPS and CNRS UMR 5589, 118 route de Narbonne, F31062 Toulouse, France*

*** LSA, Université de Lyon - UCBL and CNRS, UMR 5180, 43 Bd. du 11 novembre 1918, Villeurbanne, F-69622, France*

**Abstract**

The heat capacity of the mass selected $Na_{41}^+$ cluster has been measured using a differential nanocalorimetry method. A two-peak structure appears in the heat capacity curve of $Na_{41}^+$, whereas M. Schmidt and coworkers [M. Schmidt *et al*, Phys. Rev. Lett. **90**, 103401 (2003)] observed, within their experimental accuracy, a smooth caloric curve. They concluded from the absence of any structure that there is a second order melting transition in $Na_{41}^+$ with no particular feature such as premelting. The observed difference with the latter results is attributed to the better accuracy of our method owing to its differential character. The two structures in the heat capacity are ascribed to melting and premelting of $Na_{41}^+$. The peak at lower temperature is likely due to an anti-Mackay to Mackay solid-solid transition.

## I) Introduction

Since the first report of melting-like properties of small clusters, more than a decade ago [1,2], there have been several improvements in the range of species that can be studied and the precision of the measurements. Tin [3], gallium [4], sodium chloride [5], aluminum clusters [6] have been investigated. We report here on a detail feature of the caloric curve of $Na_{41}^+$, which is evidenced by our measurement setup and has remained undetected up to now.

The only way that has proven its efficiency so far for detecting phase transitions in clusters consists in measuring their internal energy $E(T)$, or equivalently their heat capacity $dE(T)/dT$, as a function of temperature [2]. Three different experimental methods have been developed for this purpose. They are all based on the assumption that there is a one-to-one correspondence between the internal energy and the evaporation rate of clusters. In the first method [7], evaporation is obtained by photoexciting the clusters, and the decay rate is characterized by the mean size of the photofragments. In the second experiment, the energy is brought through collisions with rare gas atoms in a drift tube, and internal energy is characterized by a fixed decay rate [8]. In the third experimental method, developed in our laboratory, the internal energy is characterized by the maximum number of atoms that can be stuck onto the clusters [9]. Experiments on positively charged sodium clusters have been carried out in both Freiburg and Toulouse, and give melting temperatures and latent heats of fusion in satisfactory agreement. Each type of experiment has advantages and drawbacks. We will emphasize in this paper one advantage of our method, due to its differential nature. The term differential means here that the heat capacity is obtained by measuring directly a differential quantity, with ΔE and ΔT small enough with respect to the variations of $\Delta E/\Delta T$. The practical way to implement the differential method consists in measuring the mean number of atoms that can be stuck onto the cluster as a function of temperature, for two different collision energies. The heat capacity is obtained from the difference of these two independent measurements. First, it ensures the validity of the finite difference approximation $\partial E(T)/\partial T \approx \Delta E/\Delta T$. Second, recording two curves allows in particular eliminating the dissociation energies from the final expression of $C(T)$, thus making the method insensitive to magic numbers [9].

The photofragmentation method consists in recording the caloric curve $E(T)$, whose derivative gives the heat capacity [7]. Basically, the main difference between Toulouse and Freiburg Methods is that in the photofragmentation method only one curve $E(T)$ is recorded and the heat capacity is deduced from the derivative of this single curve, whereas in Toulouse two curves are recorded, as shown in figure 1. Strictly speaking, the Freiburg method is also formally differential, but in this experiment $\Delta E$ (the photon energy) is much larger than the structures in $\Delta E/\Delta T$. That method works nicely as far as the relation between the number of photoevaporated atoms and the internal energy is smooth enough. This condition cannot be safely assumed when small clusters are involved : the stability of small clusters against evaporation may vary strongly from one size to the next (some particularly stable "magic" clusters require more energy than the others to evaporate [10]). It will be shown in this paper that our differential collisionnal method is insensitive to magic numbers.

We will report here the results obtained for $Na_{41}^+$. The previous study by M. Schmidt *et al* [11] described the caloric curve $E(T)$ of $Na_{41}^+$ as a continuously but smoothly increasing function, with no step or inflection. The authors concluded that there is neither first order transition nor any premelting in this cluster. Our conclusions are different. Although our experimental data are compatible with M. Schmidt's results, we will show that $Na_{41}^+$ is likely to undergo a first order melting transition, or at least a second order transition with premelting. Moreover, the caloric curve exhibits a two-peak structure that may be attributed to surface structural transition from anti-Mackay to Mackay icosahedral packing structure.

The paper is organized as follows: we will first briefly describe the experimental method for measuring the caloric curve of clusters. Our experimental results will be presented in section III. We will review some theoretical results concerning $Na_{41}^+$ in section IV that will help us discussing our experimental results (section V).

## II) Measuring caloric curves of mass selected clusters

The experimental method for extracting caloric curves from measurements of the mean number of atoms that can be attached onto a cluster has been described elsewhere [9]. In the following we will briefly describe the key points of the experimental method. Ionized sodium clusters seeded in helium are produced in a gas aggregation source and thermalized in a heat bath. They are mass selected and slowed down to a translational kinetic energy of a few eV in a special electrostatic device [12], before entering a collision chamber that contains sodium vapor. Then, attachment reaction products are mass analyzed in a time of flight mass spectrometer. The number of atoms that can be attached is limited by the evaporation of the clusters, which are heated by each sticking collision. The evaporation rate, thus the mean number of attachments is directly related to the initial internal energy of the clusters $E_0$ before they enter the collision cell. It is also related to the collision energy brought at each attachment. Increasing the initial temperature by $\Delta T$ will increase evaporation rate, and consequently decrease the mean size at the output of the cell. This change in the mean size can be compensated by decreasing the collision energy $E_c$ by a known quantity $\Delta E$ in order to recover the same mean size at $(T, E_c)$ and at $(T+\Delta T, E_c-\Delta E)$. In this case, the collisional energy decrease $\Delta E$ has exactly compensated the initial internal energy increase $C(T)\Delta T$, where $C(T)$ is the heat capacity at temperature $T$. In the experiment, we rather record the average size $\bar{n}$ (or equivalently the average number of stuck atoms, as shown in figure 1) of the mass distribution as a function of $T$ for a given collision energy. Two such curves are plotted in figure 1, corresponding to two different collision energies $E_{c1}$ and $E_{c2}$. The difference in the internal energy $\Delta E$ corresponding to the temperature difference $\Delta T$ in figure 1 is roughly equal to

$$\Delta E = \bar{n}\left(E_{c1} - E_{c2}\right) \qquad (1)$$

so that the heat capacity (for $T$ at the middle of the interval $\Delta T$ ) is

$$C(T) = \frac{\bar{n}\left(E_{c1} - E_{c2}\right)}{\Delta T} \qquad (2)$$

$\Delta T$ and $\Delta E$ can be set as small as required, within the experimental accuracy of their measurement. $\Delta E$ is generally of the order $10^{-1}$ eV. This energy variation is small enough to apply to a good accuracy the finite difference approximation $\partial E(T)/\partial T \approx \Delta E/\Delta T$ .

This is a simplified presentation of the principle of the method. It can be shown that this kind of method provides rigorously the heat capacity $C(T)$, without any particular assumption, such as the relation between evaporation rate and internal energy or even the dissociation channels (monomers, dimers) [9]. In particular, measuring the difference between two curves allows eliminating all the dissociation energies in the equations. Thus, our method is insensitive to the variations of dissociations energies and remains reliable even when magic numbers are sampled. Furthermore the effects of evaporation and collision statistics, energies distribution, etc, have been carefully analyzed by Monte Carlo simulations. More details and a usually unnoticeable correction to equation (2) can be found in reference [9].

## III) Experimental results on $Na_{41}^{+}$

Temperature-dependent heat capacity of $Na_{41}^{+}$ is shown in figure 2 and compared to the data from the Freiburg's photofragmentation experiment of M. Schmidt *et al* [11]. A curve averaged over several differential experiments is displayed in figure 3.

Our results are compatible with the ones of M. Schmidt *et al* [11]: we find the same absolute value and the range of variation of the heat capacity has the same magnitude. However, contrary to the conclusions of M. Schmidt *et al*, we are able to identify peaks in the heat capacity. The origin of the peaks will be discussed later, but we can already suppose that we observe a first order phase transition. In what follows, we will first analyze our results and the ones of M. Schmidt *et al* in order to understand the origin of the discrepancy.

If one supposes that the variation of $\bar{n}$ with the initial internal energy is linear, the quantity $d\bar{n}/dT$ is proportional to the heat capacity. It is easy to show that linearity occurs if all the dissociation energies of the involved sizes are equal. On the other hand, if dissociation energies are unequal, the proportionality is not true, but this leads to accidents in the curve for a given mean size $\bar{n}$ rather than for a given temperature. This point is illustrated in figure 4.

We have therefore a criterion to discriminate between accidents due to peaks in the heat capacity and accidents due to the variation of the dissociation energy for different sizes: in the former case, the accidents are at the same temperature for two different collision energies, while in the latter, they occur at the same mean size, that is with a shift of temperature equal to $\Delta T$, the horizontal distance between the two curves in figure 1. We plot in figure 5 the derivatives $d\bar{n}/dT$ corresponding to fig. 1, without shift and with a shift of $\Delta T = 17K$. Clearly, the peaks occur at the same temperature, confirming the validity of the measurement. The differential method we use allows to increase further the accuracy, to the expense of introducing more noise in the data. This noise can be averaged out by performing several experiments.

The Freiburg's group method is based on a direct measurement of $d\bar{n}/dT$ ($\bar{n}$ is rather the average number of evaporated atoms in this case [13]). In that method, one takes advantage of averaging the evaporation rate over a set of clusters of different size range, which come from the absorption of a different number of photons by the parent cluster[13].However, for small clusters where the evaporation rates of clusters may vary strongly with the size, this method does not work safely anymore. A way to get rid of this uncertainty is to vary photon energy, in order to probe clusters of different sizes: If the result does not depend on the photon energy, the caloric curve is likely to be correctly determined. Nevertheless, if it is not the case, the result must be discarded, there is no way to eliminate the problem and the heat capacity cannot be measured.

The differential method is, theoretically, definitely better than any non-differential method, since it is really able to get rid of artifacts due to the high stability against evaporation at "magic" sizes. However, as seen in figure 3 and figure 5, measuring differences between two noisy curves close to each other does not provide an excellent signal to noise ratio, which is the main drawback of the method. However, the result obtained using the non differential data processing described above may, in some cases, be a useful tool to probe the differential measurements.

Concerning $Na_{41}^{+}$, reproducible experiments have shown the structure of the caloric curve displayed in figure 2. This two-peak structure could not be reliably deduced from previous experiments, based on the measurement of the evaporation rate of small clusters, in a size range where magic numbers [14] dramatically perturb the measurements. Caloric curves with several peaks have already been observed for aluminum clusters; these multiple peaks have been ascribed either to uncompleted thermalization or to premelting [15,16,17,18]. Fine structures in the caloric curves of $Na_{139}^{+}$ and $Na_{147}^{+}$ also revealed special premelting features, but the caloric curves do not exhibit multiple peaks but only weak shoulders [19]. A peak before the "genuine" melting transition can be due either to a surface premelting or to a solid-solid-structural transition. There is no experimental way to distinguish between these two phenomena so that one has to rely on numerical simulations.

## IV) Theoretical caloric curve of $Na_{41}^{+}$

S. M. Ghazi and coworkers [29] have already calculated heat capacity curves of sodium clusters for n=39, 40 and 41, for positive, neutral and negative charge states. In their calculations, the main maximum of the caloric curve of $Na_{41}^{+}$ peaks at about 305 K. At lower temperatures, roughly around 250 K, there is a weak shoulder in the curve, which also appears for other clusters, particularly for $Na_{40}^{+}$. Unfortunately, the origin of these shoulders is not analyzed in this paper.

The calculations presented in figure 3 were carried out in the same way as in reference [20], where F. Calvo and F. Spiegelmann have analyzed the structural changes in the melting transition of sodium clusters. In the vicinity of the size n=40, using both a classical empirical potential (SMA) and a quantum tight-binding model (TB), they found a bump in the caloric curve before the main melting peak, which was attributed to a competition between polyicosahedral (anti-Mackay) and multilayer icosahedral (Mackay) structures. Additional simulations were carried out using the same potential energy surfaces, employing the parallel tempering strategy for improved convergence [21]. The stable structures of $Na_{41}^{+}$ predicted by the two potentials also have a polyicosahedral character, in agreement with the work of Ref. [22] on similar clusters. The caloric curves, also corrected for vibrational delocalization using the Pitzer-Gwinn approximation [23,24], are shown in figure 3 together with the present experimental measurements. Most of the quantitative differences have been discussed previously [20], and essentially consist in a systematic overestimation of the melting temperatures by the TB model with respect to the SMA model. Nevertheless, some interesting qualitative similarities are found below the melting peak under the form of clear premelting shoulders near 120 and 60 K in the TB and SMA models, respectively. These features are essentially washed out when taking quantum nuclear effects into account, but they remain visible for the more realistic TB calculation.

**V) Discussion**

The variations of the melting temperatures of sodium clusters have been interpreted by geometrical arguments: high melting temperatures are related to high stability of geometric magic numbers [25,26]. However, although geometric structure changes seem to be the major reason for the variations of melting temperatures of clusters, it has been experimentally demonstrated, in the case of aluminum clusters, that electronic effects also play a role [27]. The effects of electronic structure [28] and charged state of sodium clusters have been theoretically addressed by [29]. The existence of melting transitions characterized by multiple peaks in the caloric curve had already been observed and commented in the case of aluminum clusters. Well resolved peaks in heat capacities may result from two possible causes if only ground state isomers are assumed to be produced [16]: surface melting, or structural transition. Multiple Peaks or dips may also be due to the presence of metastable clusters formed before the thermalization stage [15,30]. Multiple peaks in the heat capacity had never been observed in sodium clusters (nor in any clusters other than aluminum clusters to our knowledge) although they had been theoretically predicted. This paper presents for the first time evidences of such a behavior in sodium.

Our experimental caloric curve for $Na_{41}^+$ is not in quantitative agreement with the theoretical predictions of references [21] and [29]. Nevertheless, similarities between experiment and calculations can be pointed out. First, a phase transition with premelting can be identified in $Na_{41}^+$. It is not easy, within experimental accuracy, to decide whether this is a first or second order phase transition: There is no firm evidence that the second structure is really a peak with a very small latent heat, rather than merely a shoulder. Whatever the conclusion about this last point, M. Schmidt *et al* arrived to a different conclusion, namely that the heat capacity of $Na_{41}^+$ continuously increases from solid to liquid with no peak in between, not even a smoothed one [31]. This kind of behavior was not so unexpected since second-order phase transitions have already been observed in gallium clusters [32]. Theoretical predictions also show that small sodium clusters may melt gradually [33]. Second, the temperature gap between the melting temperatures and the shoulders is of the same order of magnitude in the experiment and in the calculations, even though in references [21] and [29] the structural solid-solid transition produces only a shoulder and not a well resolved peak in the caloric curve of $Na_{41}^{+}$.

## Conclusion

The heat capacity of the free cluster $Na_{41}^{+}$ has been measured as a function of temperature. Two well resolved peaks have been identified. The simulations presented in this paper (together with the ones of reference [20]) suggest that the origin of the double peak structure in the experimental caloric curve of $Na_{41}^{+}$ is likely to originate from a surface structural transition from anti-Mackay to Mackay structures.

## Acknowledgements

The authors are greatly indebted to Florent Calvo for the calculations shown in figure 3. We are also particularly grateful to Martin Schmidt for his invaluable help to assess and compare the Freiburg and Toulouse methods.

**Figure captions**

**Figure 1** : Evolution of the mean size $\bar{n}$ of the mass distribution after attachment of sodium atoms onto $Na_{41}^{+}$ as a function of the initial temperature of the cluster ($\bar{n}$ is labeled here respective to the initial mass). Opens circles are raw data and the lines show the smoothed data (low pass FFT filter) used to determine the heat capacities. The two curves have been recorded for two different collision energies : $E_{c1}$ = 240 meV (upper curve, red online) and $E_{c2}$ =480 meV (lower curve, blue online) in the center of mass frame. The horizontal gap $\Delta$ between the two curves is around 17K.

**Figure 2** : Heat capacity of $Na_{41}^{+}$ as a function of temperature. The results obtained from three separate experimental runs are shown, corresponding to different collision energy pair ({0.24 and 0.37 eV}, {0.24 and 0.49 eV} and {1.22 and 1.34 eV} in the center of mass frame). They are compared to the results obtained by M. Schmidt and coworkers (large open circles, red online) [11,34].

**Figure 3** : Solid line (red online): experimental heat capacity of $Na_{41}^{+}$ averaged over the 3 curves of figure 2. Dashed-dotted [dashed] line : heat capacity extracted from Monte Carlo simulations using the SMA [Tight Binding] model with quantum vibration [21]. The SMA model with classical vibration (dotted line) exhibits more clearly the bump (arrow).

**Figure 4** :

*Top :* Schematics example of the evolution with temperature of the barycenter $\bar{n}$ of the mass distribution, shown at two collision energies $E_{c1}$ (dotted line, blue online) and $E_{c2}$ (dashed-dotted line, red online). A genuine variation of heat capacity (labeled *1*) produces two curves locally self-similar through a *vertical* shift, whereas an artifact (labeled *2*), due to a variation with the size of the stability against evaporation, produces two curves locally self-similar through an *horizontal* shift.

*Bottom :* Top and middle curves $C(E_{c1})$ and $C(E_{c2})$ are erroneous two-peak heat capacities deduced from the derivatives $d\bar{n}/dT$ of the two curves above (non differential method). The bottom curve $C(T)$ is the correct heat capacity obtained from the variation of $1/\Delta T$ along the two curves (differential method).

**Figure 5** : Derivatives of the two curves shown in figure 1. Thin (red online) line : $E_c$=240meV. Thick (blue online) line : $E_c$=480 meV. a) both derivatives are represented using the same temperature scale. b) The temperature scale of the curve at $E_c$=240 meV is shifted by 17 K. The better overlap in figure a) proves that a particular stability of some masses is not responsible for the two-peak structure of $d\bar{n}/dT$ (see text).

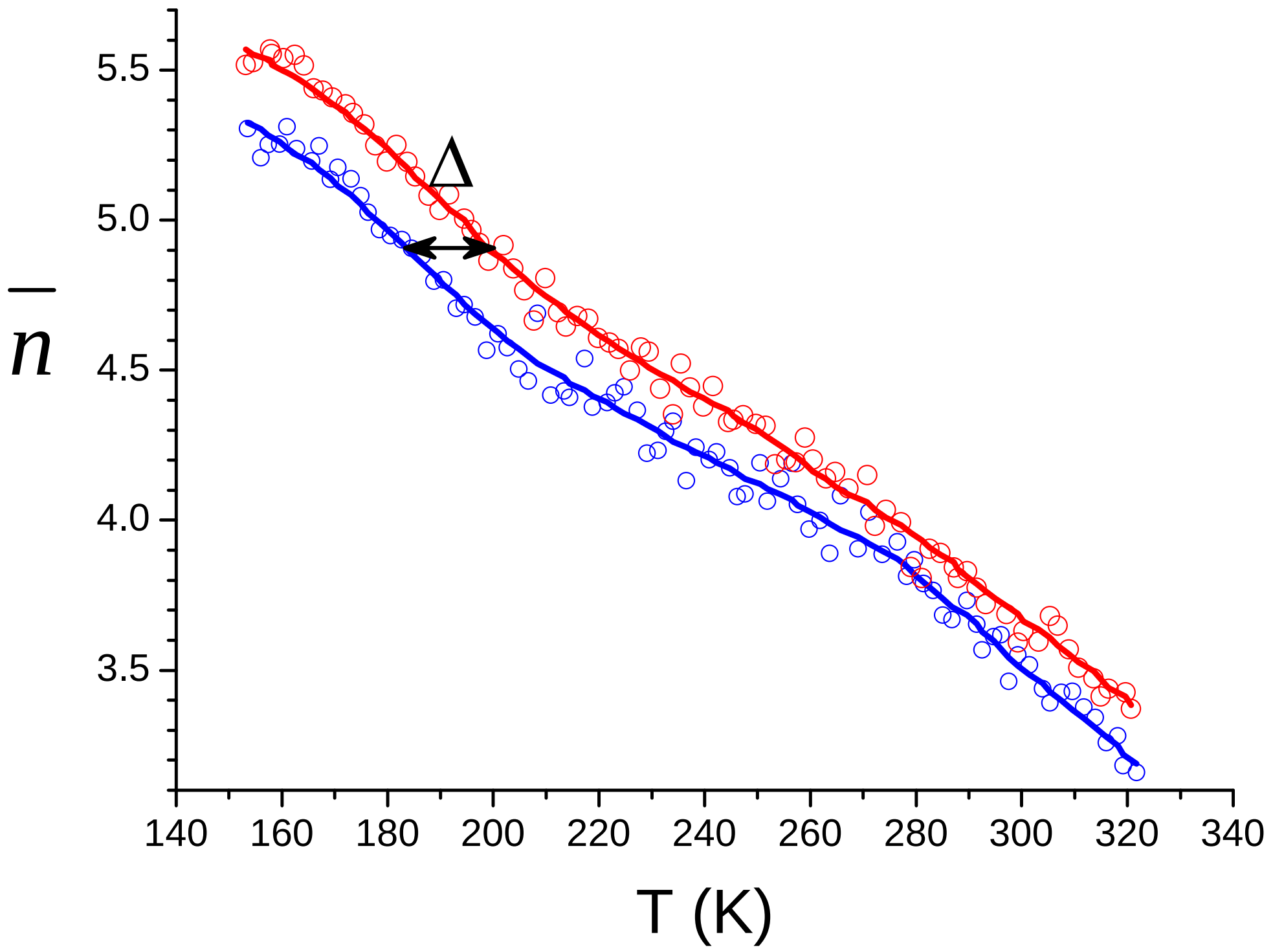

**Figure 1**

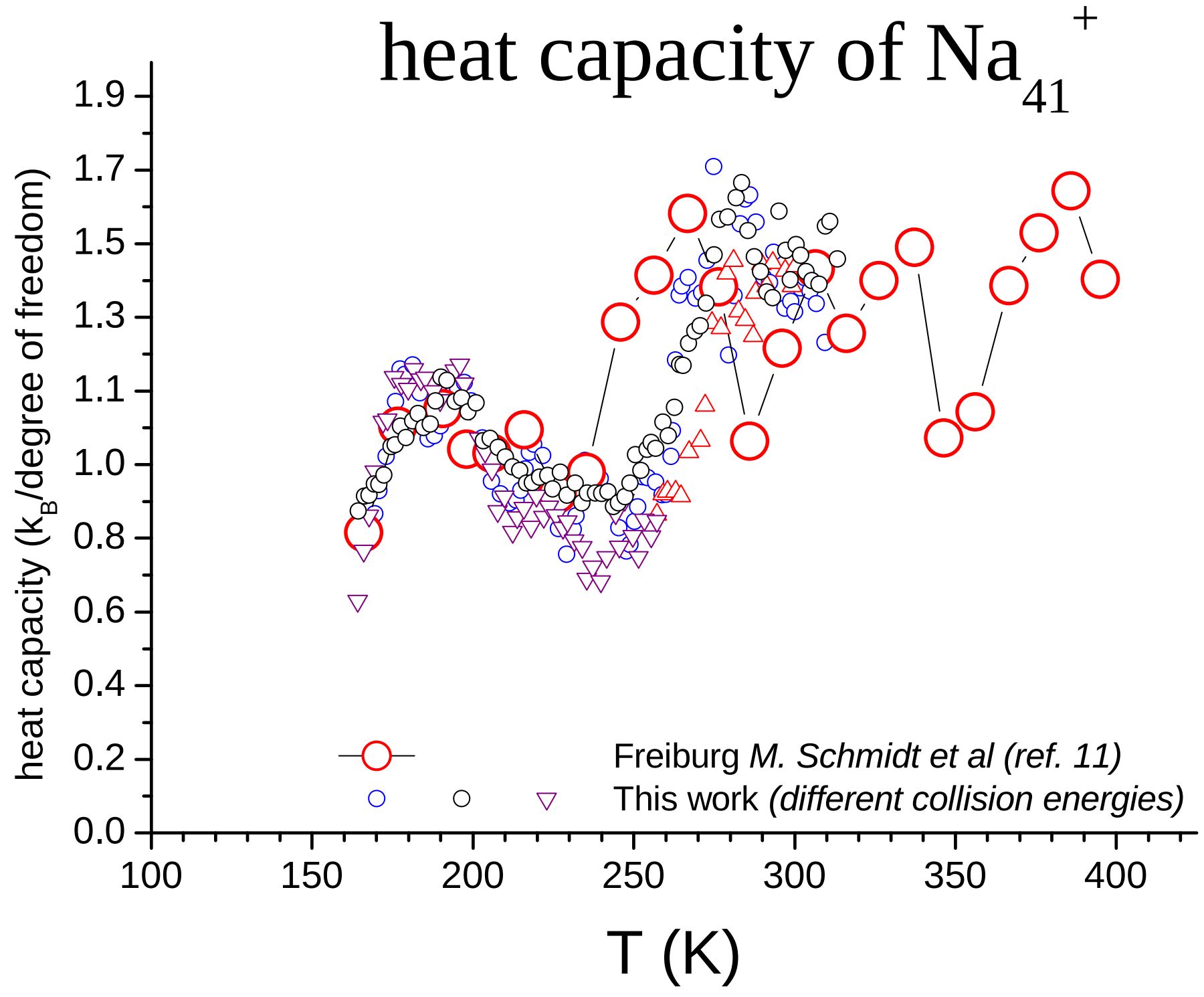

**Figure 2**

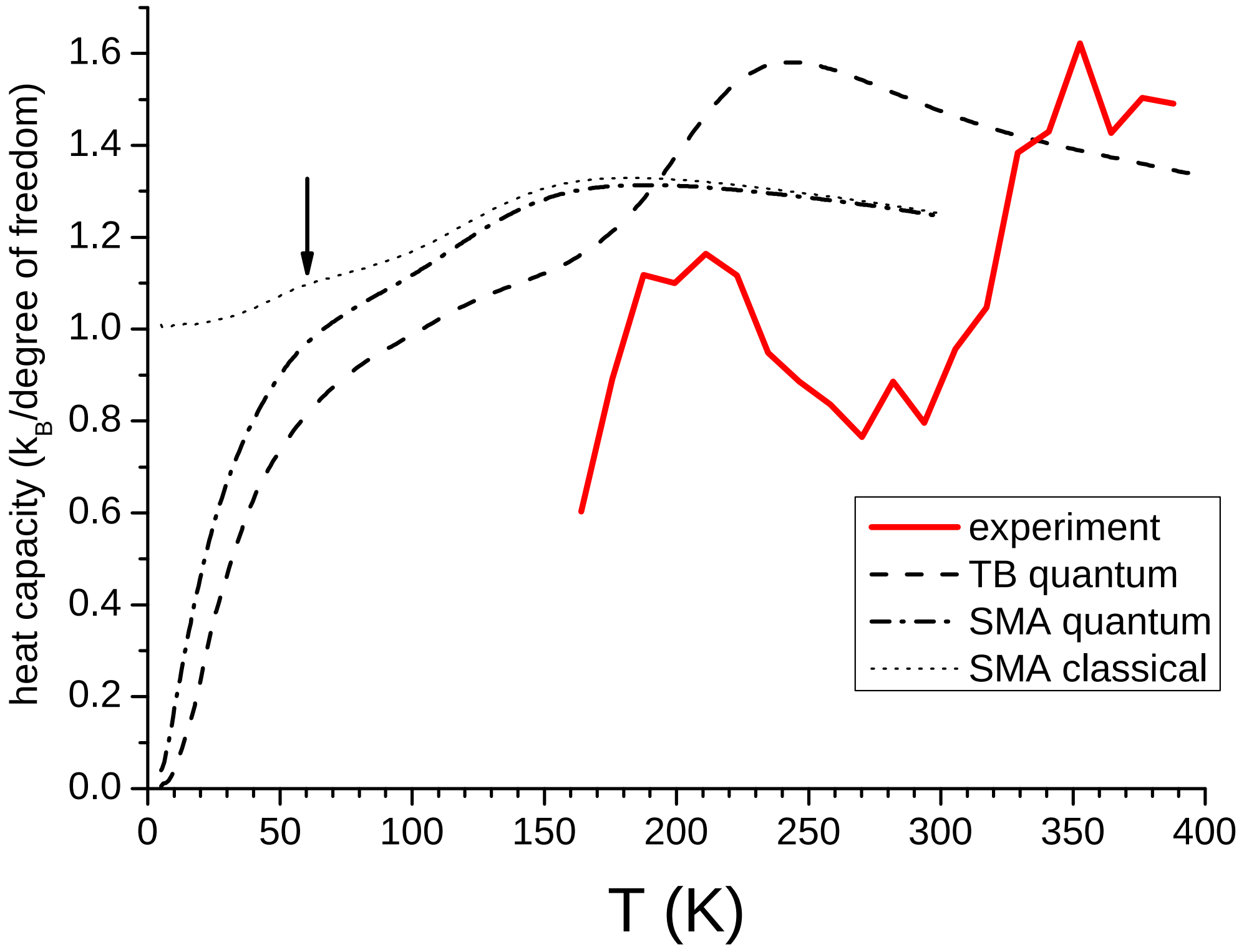

**Figure 3**

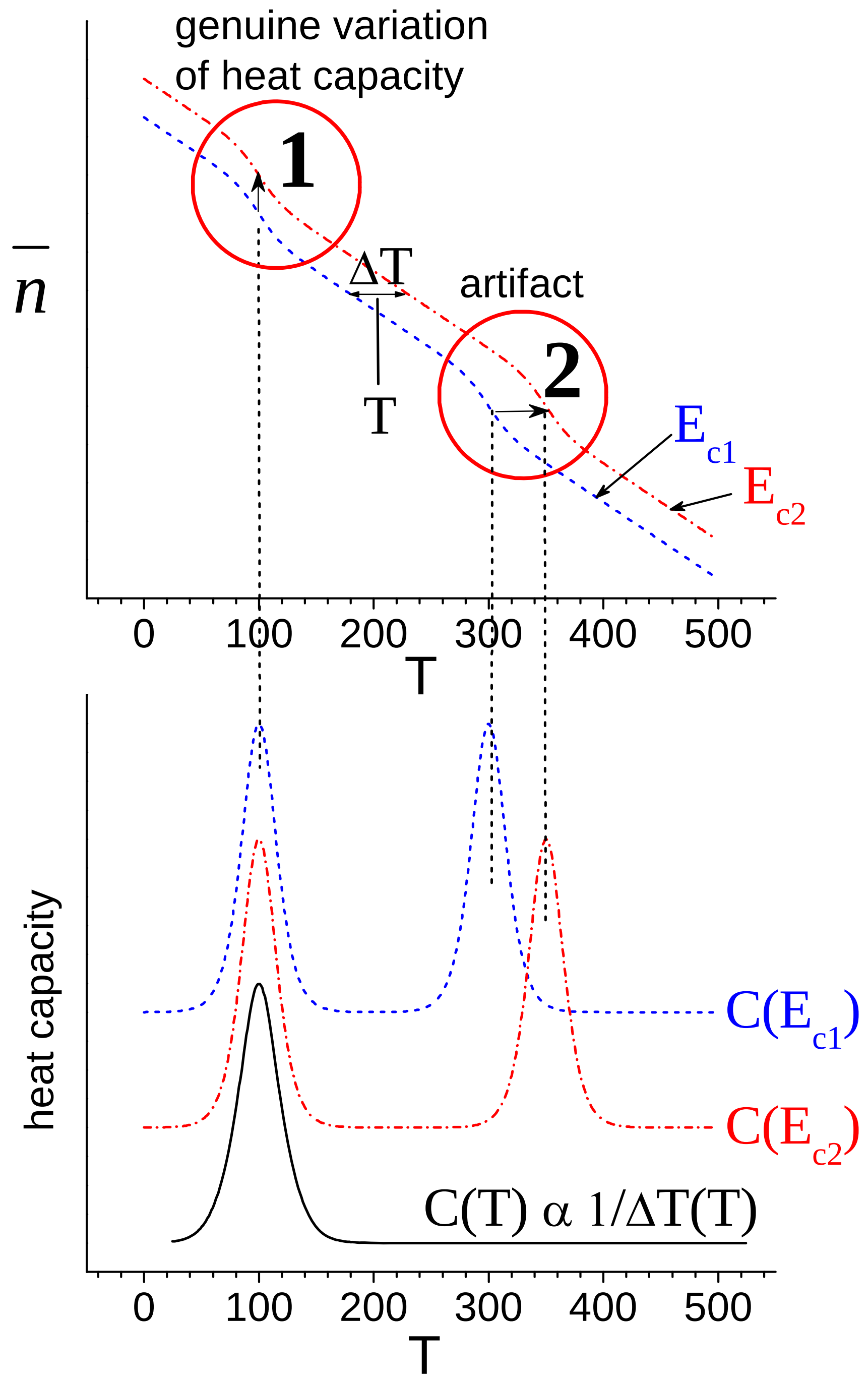

**Figure 4**

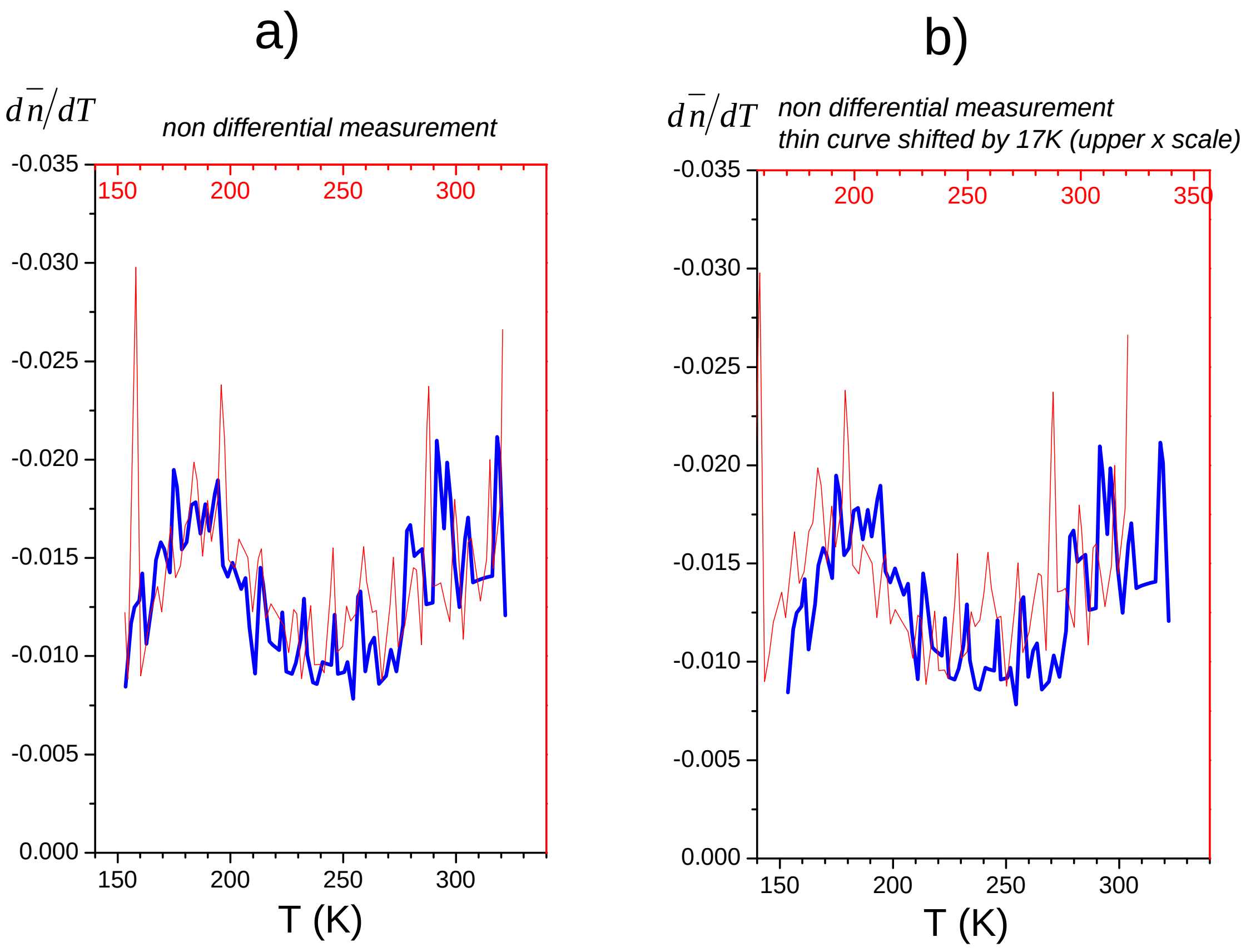
a)
$d\bar{n}/dT$
non differential measurement
-0.035
-0.030
-0.025
-0.020
-0.015
-0.010
-0.005
0.000
150
200
250
300
T (K)
b)
$d\bar{n}/dT$
non differential measurement
thin curve shifted by 17K (upper x scale)
350

**Figure 5**